# Pattern stabilization through parameter alternation in a nonlinear optical system

J. P. Sharpe[1*], P. L. Ramazza[2], N. Sungar[1] and Karl Saunders[1]

[1]Dept. Physics, Cal Poly State University, San Luis Obispo, CA 93407, USA

[2]Istituto Nazionale di Ottica, Largo E. Fermi 6, I50125 Florence, Italy

Abstract:

We report the first experimental realization of pattern formation in a spatially extended nonlinear system when the system is alternated between two states, neither of which exhibits patterning. Dynamical equations modeling the system are used for both numerical simulations and a weakly nonlinear analysis of the patterned states. The simulations show excellent agreement with the experiment. The nonlinear analysis provides an explanation of the patterning under alternation and accurately predicts both the observed dependence of the patterning on the frequency of alternation, and the measured spatial frequencies of the patterns.

[*] jsharpe@calpoly.edu

Recent theoretical and numerical studies have drawn attention to the patterns that can arise when spatially extended nonlinear systems are driven by a global periodic modulation of the system parameters[1-3]. In [2] a Turing-type reaction-diffusion system is shown to form patterns due to short-time instabilities, a mechanism that is applicable only to multi-field systems. In [3] the single-field Swift-Hohenberg model is investigated as the system is switched between two values of a parameter, each of which individually leads to a homogeneous state under steady state conditions. In this case the mechanism derives from an effective averaging of otherwise monostable potentials to produce a bistable potential under which patterning occurs. The theoretical analysis of [3] obtains mode amplitude equations for the one-dimensional case while the structure of the Swift-Hohenberg model then permits an inversion symmetry argument (without derivation of mode amplitude equations) to infer the type of patterns in the two-dimensional case. Numerical simulations confirm this.

In this Letter we report the first experimental observation of pattern stabilization in a nonlinear system when one of the control parameters is alternated between two values, neither of which individually leads to a patterned state. The pattern forming mechanism is related to the Swift-Hohenberg model described in [3] but, as will be discussed below, has several fundamental differences. We have also carried out a comprehensive theoretical analysis, including a derivation of the two-dimensional mode amplitude equations which we have used to predict the length scales and types of pattern that can arise. These are in good agreement with the observations.

The experimental arrangement we use is that of a liquid crystal light valve (LCLV) placed in an optical feedback loop. This system exhibits a wide range of spatiotemporal behavior and a comprehensive review can be found in [4]. The LCLV is an optoelectronic device composed of thin layers of liquid crystal and amorphous silicon across which is placed an alternating voltage. Light of spatially varying intensity incident on the amorphous silicon ("write") side of the device causes a varying electric field across the liquid crystal. A light beam incident on the liquid crystal ("read") side of the device passes through the liquid crystal, reflects from a dielectric mirror which serves to isolate the read and write sides, and emerges with a spatially variant phase modulation imposed on it. The transverse dimension of the device is ~ 1cm$^2$. Further details of the particular device used here can be found in [5].

The function $f(I)$ that relates the change in phase of the read light to the intensity of light at the write side of the LCLV depends on the voltage and frequency used to drive the device. For an 8V amplitude, zero offset, 3kHz drive the measured phase response is shown in the inset of figure 1. This phase response is empirically well fitted by the equation

$$f(I) = 10.5 \tanh(2.5I) \qquad (1)$$

and is also plotted. With the above driving parameters the phase retardation at zero input intensity was determined to be $\phi_0 = \pi$ rad.

The experimental setup is shown in figure 1. Linearly polarized light from an argon ion laser (λ=514 nm) was intensity modulated by using the first diffraction order from an acousto-optic modulator. The polarization orientation of the light was set by the half wave plate and, after spatial filtering, the beam was allowed to fall on the liquid crystal side of the LCLV. Upon reflection the beam was routed via lenses L1, L2, and the imaging fiber bundle onto the amorphous silicon side of the LCLV. The various distances between the optical components are as shown in the figure caption and lead to an overall self-focusing nonlinearity [6]. For the measurements reported below, the incoming light was polarized at 45º to the liquid crystal director and the polarizer (P) set with its transmission axis also at 45º. The pinhole P1 was adjusted to admit only the first two critical q-vectors of the spatial instability. With these settings one can find many different spatiotemporal behaviors including chaotic patterning and localized structures, depending on the input light intensity [5].

In this setup the equation governing the evolution of the phase response of the LCLV as a function of the input light is given by [6]

$$\tau \frac{\partial \phi(\vec{r},t)}{\partial t} = -(\phi(\vec{r},t) - \phi_o) + l_d^2 \nabla_\perp^2 \phi(\vec{r},t) + f(I_{fb}) \qquad (2)$$

where $\tau$ is the relaxation time of the device (in this case about 30 ms), $\phi_o$ is the constant-phase retardation of the LCLV with no write light, $l_d$ is the (transverse) diffusion length of the LCLV (about 50 µm) and the function $f$ relates the phase modulation induced in the liquid crystal to the light that is fed back on the LCLV (equation 1) with intensity $I_{fb}$.

With a polarizer inserted in the optical loop the feedback intensity is given by

$$I_{fb} = I_o \left| \exp\left(\frac{-iL\nabla^2}{2k_o}\right) \times (Be^{-i\phi} + C) \right|^2 \qquad (3)$$

where $I_o$ is the input intensity of the laser beam, $L$ is the free-space propagation length in the feedback loop and $k_o = 2\pi/\lambda$. $B = \cos(\theta_1)\cos(\theta_2)$ and $C = \sin(\theta_1)\sin(\theta_2)$ where $\theta_1$ and $\theta_2$ are the respective angles between the input light polarization and the polarizer transmission axis with the liquid crystal director.

In order to explore the effect of alternation the input light intensity to the system was switched between high and low values at unit mark-space ratio. It was found that, depending on the values of the high and low intensities and the frequency of modulation, a number of effects could be observed on the formation of patterns. One of the most interesting, which we will focus on in this report, was observed when the light intensity was modulated between 3.9 and 0.1 mW/cm². (This was measured at the write side of the LCLV. All power measurements reported here have an uncertainty of about five percent). At constant intensities of 3.9 mW/cm² or 0.1 mW/cm² the output of the system was homogeneous (as shown in figures 2 (a) and (b)). At intermediate constant intensities we observed formation of irregular rolls and the appearance of bright regions at intensities above ~ 2 mW/cm² (figure 2(c)). The patterns had spatial frequencies corresponding to $\hat{q}^2$ from about 1.5 to 4 where we define the non-dimensional spatial frequency as $\hat{q} = \sqrt{\dfrac{Lq^2}{2k_0}}$ with $q$ the wave vector of the pattern. The bright regions grow as the intensity increases and above ~3 mW/cm$^2$ the system settles into the homogeneous bright state.

When a periodic modulation was applied to the input light intensity at a frequency of some 6 to 20 Hz (periods comparable to the relaxation time of the device), hexagonal patterning was seen. When the modulation is slow enough ( < 2 Hz) then the system alternates between the two homogenous states and figure 2(d) shows the system output when the intensity is high. At 14 Hz (figure 2(e)) the strongest hexagonal patterning occurred, as judged by eye on the CCD monitor. These hexagons had $\hat{q}^2 = 2.8 \pm 0.2$. As the alternation frequency was raised beyond some 20 Hz (figure 2(g)) the pattern became more disordered and showed rolls as well as hexagons. At frequencies beyond a hundred hertz (figure 2(h)) the pattern was similar to that seen with constant intensity input set at the mean value of the alternation intensities (2mW/cm$^2$).

The behavior of the optical system was simulated by numerically integrating equation (2) and using equations (1) and (3). The simulations are in good quantitative agreement with the experimental results. In particular, they show that there is a narrow range of alternation frequencies of order 10 Hz for which there is stable hexagonal pattern formation. We estimated the spatial frequencies in the pattern by computing the power spectral density of the pattern just after the intensity of the input goes high and integrating the spectrum in the azimuthal direction. For a modulation frequency of 14 Hz this gave a peak corresponding to $\hat{q}^2 = 2.9$ with a peak FWHM of 0.4. Outside of this frequency range there is either alternation between homogeneous states at lower frequencies or the formation of irregular patterning at higher frequencies.

Insight into the system can be gained by considering the homogeneous solution of equation 2. The fixed points $\phi^*$ for the homogeneous solution as a function of input intensity $I_0$ are shown in Fig. 3. A combination of a linear stability analysis and simulation show that a homogeneous stable state exists for $I_0 < 0.2$ mW/cm$^2$ with $\phi^* = \pi$, and for $I_0 > 3$ mW/cm$^2$ with $\phi^* = \pi + 10.5$. These results are consistent with the dark and bright states observed for the lower (<0.11mW/cm$^2$) and upper (> 3mW/cm$^2$) intensities, respectively. In the intermediate range of intensities the patterns observed in experiment and from simulations are also indicated in figure 3.

Computer simulations show that the observed patterning (both for fixed and alternating intensities) takes place about the middle branch of figure 3. To mathematically analyze the behavior of these patterns we apply a patterned perturbation about the fixed point and expand the phase response equation to third order, following the basic scheme of [7]. Three Ginzburg-Landau type mode amplitude equations are derived with coefficients obtained using the computer algebra package MAPLE. These are too long to show here. The implicit time dependence of the coefficients (through their dependence on $\phi(t)$) plays an important role in studying the behavior under alternation of intensity.

The results of a stability analysis of these amplitude equations are summarized in the state diagram of figure 4 where the horizontal axis shows the stable fixed point phases that correspond to a given input light intensity, also shown in the figure. From the diagram one can see that there are no stable hexagons with $\hat{q}^2 \sim 3$ at either the high intensity of 3.9 mW/cm$^2$ (corresponding to a phase of 8.9 rad) or at low intensity of 0.11 mW/cm$^2$ (corresponding to 5.6 rad, where the middle branch appears). The diagram indicates that at intermediate fixed intensities hexagons with $\hat{q}^2 \sim 3$ could be stable though we find in actuality that the system selects irregular roll patterns at the lower value $\hat{q}^2 \approx 2$.

At higher intensities (> 3mW/cm$^2$) both patterned and homogeneous states are possible but the system eventually settles into the homogeneous bright state. This can be understood analytically by the fact that, even though they are negative, the eigenvalues for stabilization against perturbations away from roll and hexagon states approach zero as $I_0$ is increased, while the corresponding negative eigenvalue for the homogeneous state remains finite.

To study the effects of alternation of the input intensity we numerically integrated the amplitude equations taking into account that the homogeneous phase moves along the middle branch. First, we imposed small amplitude hexagonal or roll patterns about the phase corresponding to the high intensity and then computed the amplitudes of the patterns as a function of time after the intensity was set to 0.1 mW/cm$^2$. Then we imposed small amplitude hexagonal or roll patterns about the phase corresponding to the low intensity, set the intensity to 3.9 mW/cm$^2$, and again computed the amplitudes of the patterns as a function of time. The results of these calculations for hexagons are shown in figure 5 for times up to the relaxation time $\tau$ (corresponding to the duration of high and low intensities at frequencies producing strong patterning). Rolls lead to very similar results and are not shown. Consider first when the intensity is switched from low to high (Figure 5(a)). For both hexagons and rolls the amplitudes for all values of $\hat{q}^2$ decrease when the time $t < \tau/2$. This is then followed by a rapid increase in amplitude for $2 \leq \hat{q}^2 \leq 4$ and decay for other values of $\hat{q}^2$. We note that the fastest growing mode is around $\hat{q}^2 = 2.7$. Figure 5(b) shows the case when the intensity switches from high to low. In this case, for both hexagons and rolls, the modes corresponding to $2 \leq \hat{q}^2 \leq 3$ increase, with the largest rate also around $\hat{q}^2 = 2.7$ for $t<0.8\tau$. For all the other values of $\hat{q}^2$ there is a decrease in the amplitudes. These results agree well with computer simulations of the system. We would thus expect that as the intensity is switched up and down with a period of order $2\tau$ the patterns with $\hat{q}^2 \sim 3$ will dominate, as we observe.

In order to explain the selection of hexagons over rolls we refer to calculations leading to the state map (Fig. 4). Using our simulations we first note that at the optimum frequency of alternation the homogeneous phase varies between 6.5 - 9 radians. As can be seen from figure 4 in the upper regime (around a phase of 9 radians) rolls with $2 \leq \hat{q}^2 \leq 3$ are stable. In this range, although static solutions for hexagons exist, they are unstable to small perturbations. However, both the negative eigenvalues for rolls and the positive eigenvalues for hexagons are very small indicating a slow growth or decay of the instabilities. In the lower regime (around a phase of 6.5 radians) hexagons with $2 \leq \hat{q}^2 \leq 3$ are stable. Roll solutions with $2.7 < \hat{q}^2 \leq 3$ do not exist and for $2 \leq \hat{q}^2 \leq 2.7$ they have very large positive eigenvalues indicating a rapid growth of instabilities. Therefore we would expect hexagons to be strongly favored over rolls in the lower regime while in the higher regime, although the hexagons are unstable, the growth rate of the instability is small enough to limit pattern decay.

A plausible qualitative mechanism for our observations of hexagonal patterning can be summarized as follows. In the high intensity regime the low spatial frequency modes are suppressed due to their fast decay while in the low intensity regime the high spatial frequency modes are suppressed for the same reason. Then, the time of alternation is such that it is long enough to permit the growth of the intermediate spatial frequencies but short enough to prevent the development of either of the homogeneous states. Finally, in this particular case, hexagons are selected over rolls due to the greater instability of rolls at the low intensity.

In conclusion, we have presented an experimental realization of an extended nonlinear dynamical system that exhibits pattern formation as the result of alternation between two different homogeneous states. We can explain our observations on the basis of a weakly nonlinear analysis of the system which agrees quantitatively both with experiment and computer simulations. When considering this work in comparison with previous studies it seems that the nonlinear optical system most closely resembles that of [3] in the sense that the governing equations exhibit type $I_s$ instability [8], though the coupling is very different from Swift-Hohenberg. It is also of interest to note that at the high intensity the optical system is not truly monostable (see fig 3). Rather, the bright state is always selected because of the very weak stability of the other fixed points. Phenomenologically the optical system is also different. In the alternated Swift-Hohenberg model when the alternation frequency is very large an effective (i.e. just a pure average of each dynamic) bistable potential is achieved and patterning is observed. When the alternation frequency is reduced and is comparable to the relaxation time the same patterns are observed (although they are oscillating). This does not occur in the optical system. When the switching is much faster than the internal relaxation time the irregular patterns obtained correspond (trivially) to those obtained with the average parameter value. However, as the frequency is reduced, strong hexagonal patterning appears which can not be obtained with any single steady value of the parameter.

We acknowledge very helpful discussions with John Toner, University of Oregon and support of the Research Corporation, Tuscon, AZ.

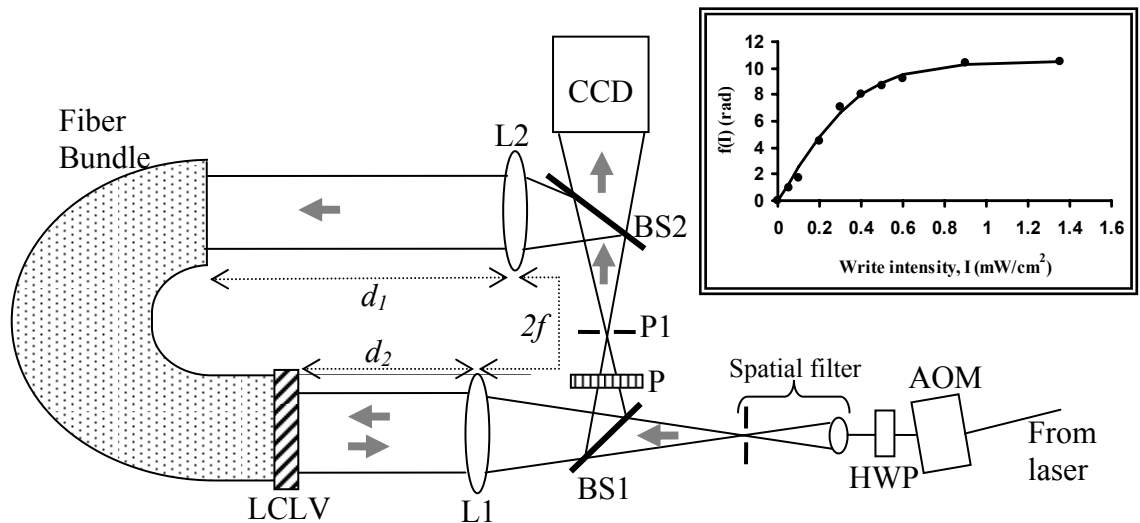

Figure 1. Schematic of the experiment. The focal length of the lenses L1 and L2 is $f=25$ cm. The distances $d_1$ and $d_2$ are 20 and 15 cm, respectively, leading to an effective propagation length of -15 cm and thus a self-focusing nonlinearity [6]. BS1, BS2 beamsplitters, P polarizer, P1 pinhole, HWP half wave plate, AOM acoustooptic modulator. Another lens and CCD camera (not shown here) are used to monitor the Fourier plane at pinhole P1. Inset: Phase response of the LCLV as a function of write light intensity and no feedback. The solid points show the experimental data while the line is $10.5\tanh(2.5I)$ where $I$ is the intensity.

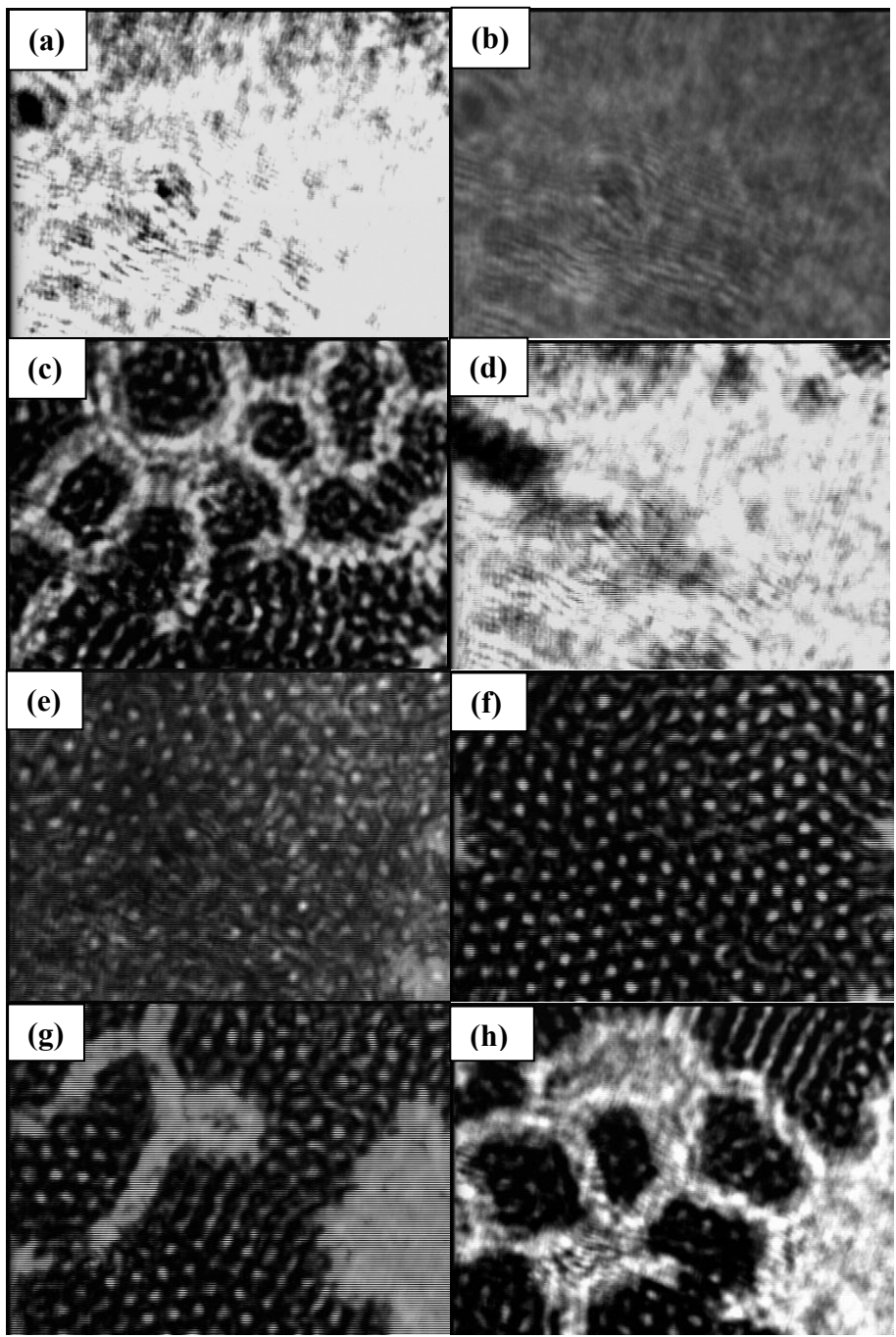

Figure 2. Output from the experiment. (a) and (b) show the homogeneous states obtained for an input intensity to the back of the LCLV of 3.9 mW/cm$^2$ and 0.1 mW/cm$^2$, respectively. (c) shows the output obtained with a constant intensity input of 2 mW/cm$^2$. (d), (e), (f), (g) and (h) show the output obtained

with the input irradiance switching between 3.9 mW/cm$^2$ and 0.1 mW/cm$^2$. The modulation frequencies are (d) 2 Hz (image obtained when the intensity goes high), (e) 6 Hz, (f) 14 Hz, (g) 20 Hz and (h) 200 Hz.

12/27/2005    16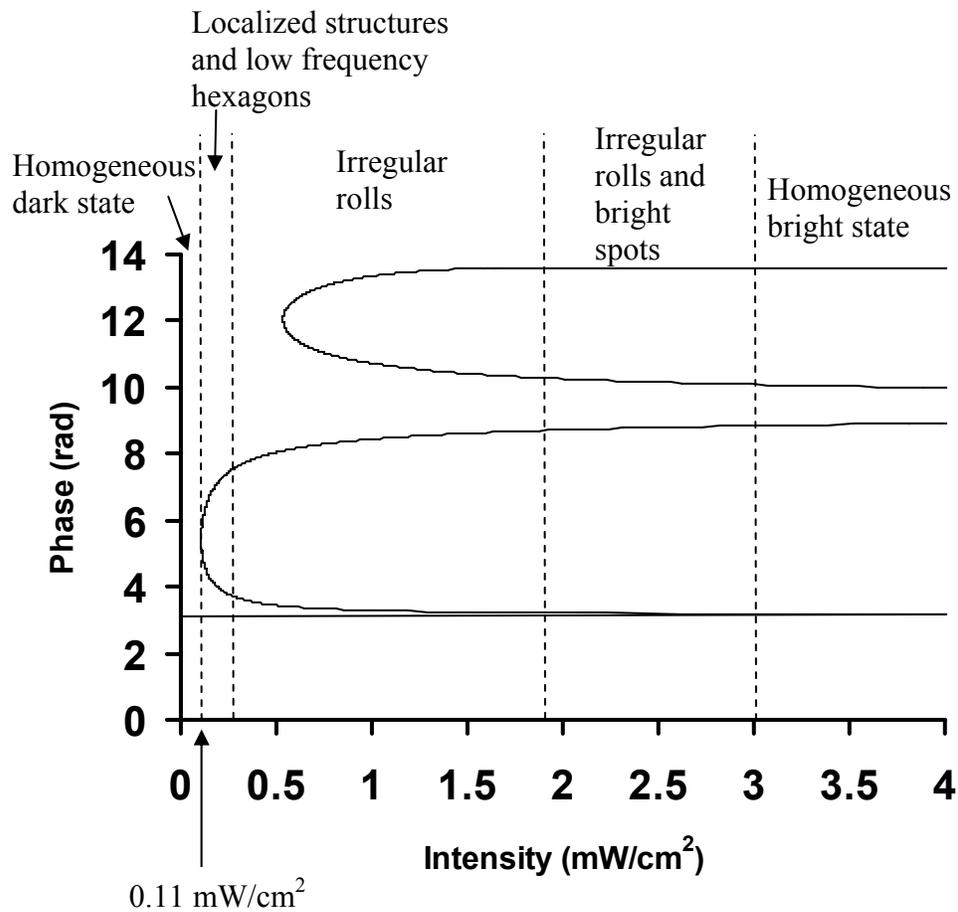

Figure 3. Homogeneous steady state phase as a function of input intensity for the LCLV feedback system with both input polarization and output polarizer oriented at 45°. Also indicated are the types of patterns seen in the experiment and in computer simulations under constant intensity illumination.

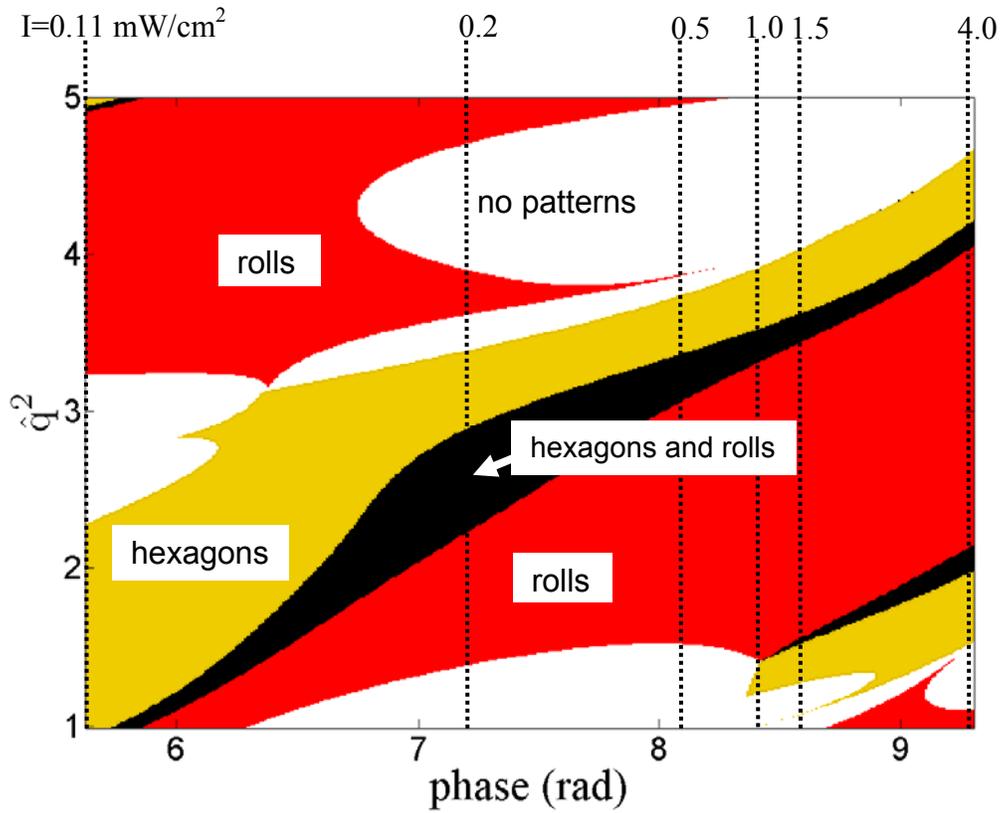

Figure 4. (color online) State map showing the regions of existence of stable patterns of rolls and hexagons as a function of spatial frequency and phase. The range of phase from about 5.5 to 9.4 is along the middle stable branch of figure 3. The corresponding intensities are also indicated on the diagram. We did not find any stable mixed states.

12/27/2005                                                                                    18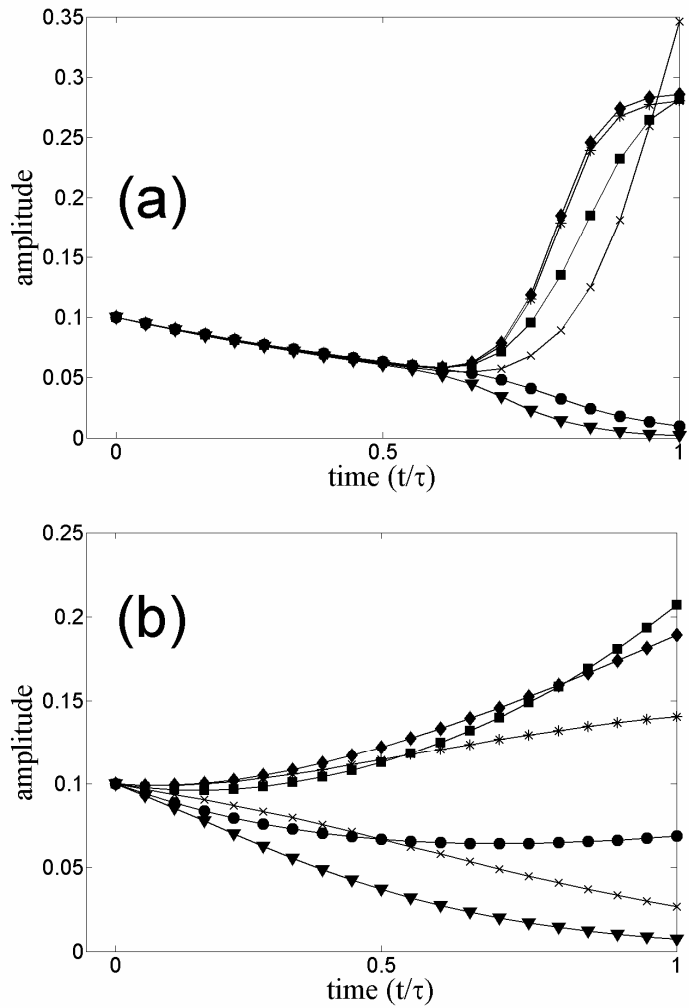

Figure 5. Amplitudes of hexagons vs. time after the intensity is switched from 0.1 to 3.9 mW/cm$^2$ (a) and after the intensity is switched from 3.9 to 0.1 mW/cm$^2$ (b). The spatial frequencies correspond to $\hat{q}^2$ = 1 (●), 2 (■), 2.7 (♦), 3 (∗), 4 (×) and 5 (▼). Symbols are to guide the eye and do not represent time steps. The amplitudes of the rolls show very similar behavior.